\documentclass[proof]{WileyASNA-v1}

\articletype{ORIGINAL ARTICLE}

\newcommand{\mcitet}[1]{\mbox{\citeauthor{#1}} (\citeyear{#1})}
\newcommand{\mciteauthor}[1]{\mbox{\citeauthor{#1}}}
\newcommand{\mcitealt}[1]{\mbox{\citeauthor{#1}} \citeyear{#1}}
\newcommand{\mcitep}[1]{(\mbox{\citeauthor{#1}} \citeyear{#1})}

\received{30 December 2025}
\revised{28 January 2026}
\accepted{30 January 2026}

\begin{document}

\title{The Bochum Survey of the Southern Galactic Disk:\break III. Complete Data Release}

\author[1]{Julia Blex}

\author[1]{Moritz Hackstein}

\author[1]{Christian Westhues}

\author[1]{Michael Ramolla}

\author[2]{\\Markus Demleitner}

\author[1]{Dominik J. Bomans}

\author[1]{Kerstin Weis}

\author[3]{Christofer Fein}

\author[1,4]{Rolf Chini}

\authormark{BLEX \textsc{et al.}}

\address[1]{\orgdiv{Astronomisches Institut}, \orgname{Ruhr-Universität Bochum}, \orgaddress{\city{Bochum}, \country{Germany}}}

\address[2]{\orgdiv{Astronomisches Rechen-Institut}, \orgname{Zentrum für Astronomie der Universität Heidelberg}, \orgaddress{\city{Heidelberg}, \country{Germany}}}

\address[3]{\orgdiv{Clavis Institut für Informationssicherheit}, \orgname{Hochschule Niederrhein}, \orgaddress{\city{Mönchengladbach}, \country{Germany}}}

\address[4]{\orgdiv{Centrum Astronomiczne im. Mikołaja Kopernika}, \orgname{Polskiej Akademii Nauk}, \orgaddress{\city{Warszawa}, \country{Poland}}}

\corres{Julia Blex, Astronomisches Institut, Ruhr-Universität Bochum,\newline Universitätsstraße 150, 44801 Bochum, Germany. \email{julia.blex@rub.de}}

\abstract{The Southern Galactic Disk Survey (GDS) monitored a mosaic of 268~fields along a $6^\circ$-wide stripe in the southern Galactic disk with simultaneous observations in $r'$ and $i'$ \mbox{($7^\mathrm{m} \lesssim r', i' \lesssim 18^\mathrm{m}$)} from September~2010 to September~2019. The survey design and data characteristics, as well as first results in $r'i'$, were presented by \mciteauthor{2012AN....333..706H} (\citeyear{2012AN....333..706H}; Paper~I). \mciteauthor{2015AN....336..590H} (\citeyear{2015AN....336..590H}; Paper~II) extended the photometry and analysis process, and introduced the first catalogue including photometry of all 268~fields in $UBVr'i'z'$ and $r'i'$ light curves comprising up to 272~observations per field made between September~2010 and May~2015. Here we describe our custom-made observational scheduler and conclude the GDS with $r'i'$ light curves of up to 407~observations per field until September~2019 and $UBVz'$ light curves for a fraction of the fields. 113\,449~distinct sources are identified as variables. Together with Paper~II, we identified 77\,592~variables that are not listed in either the International Variable Star Index (VSX) or the cross-match catalogue by \mciteauthor{2023A&A...674A..22G} (\citeyear{2023A&A...674A..22G}). All emerging catalogues, comprising light curves, photometry, and reduced images, are made publicly available via the German Astrophysical Virtual Observatory (GAVO).} 


\maketitle

\section{Introduction}
Stellar variability has been studied by dedicated time-domain surveys for several decades; among the ground-based surveys, the original All Sky Automated Survey (ASAS; \mcitealt{1997AcA....47..467P}) and the ongoing All-Sky Automated Survey for Supernovae (\mbox{ASAS-SN}; \mcitealt{2014ApJ...788...48S}), the Optical Gravitational Lensing Experiment (OGLE; \mcitealt{2008AcA....58...69U}), the Catalina Real-Time Transient Survey (CRTS; \mcitealt{2009ApJ...696..870D}), the Palomar Transient Factory (PTF; \mcitealt{2009PASP..121.1395L}), the VISTA Variables in the Via Lactea (VVV; \mcitealt{2010NewA...15..433M}), and the Zwicky Transient Facility (ZTF; \mcitealt{2019PASP..131a8002B}) are certainly worth mentioning. With regard to space-based observations, following Gaia's second data release \mcitep{2018A&A...616A...1G} and its third data release (Gaia~DR3; \mcitealt{2023A&A...674A...1G}), Gaia has become the leading survey for general stellar information, particularly in terms of distance, but also for the classification of variable sources (\mcitealt{2018A&A...618A..30H}; \mcitealt{2023A&A...674A..14R}).

The Bochum Galactic Disk Survey\footnote{https://galacticdisksurvey.space/} (GDS) is a long-term survey that has been conducted at the Rolf~Chini Cerro Murphy Observatory (OCM), formerly the Observatory Cerro Armazones (OCA), in Chile. The GDS is an excellent survey in terms of its temporal and spatial coverage of stellar sources in the Milky Way. Moreover, the GDS complements the present surveys well (see the discussion in \mcitealt{2015AN....336..590H}; Paper~II): The coverage and wavelength range of the VST Photometric H$\mathrm{\alpha}$ Survey of the Southern Galactic Plane and Bulge (VPHAS+; \mcitealt{2014MNRAS.440.2036D}), which provides single-epoch $ugri$ and H$\alpha$ photometry, and the GDS are similar, with the VPHAS+ extending to the fainter magnitude range starting from $12^\mathrm{m}$--$13^\mathrm{m}$ and the GDS covering the brighter magnitude end of \mbox{$7^\mathrm{m}$--$18^\mathrm{m}$}. The \mbox{ASAS-SN} provides $g$ and $V$~light curves with a spatial resolution of $8.0'' / \mathrm{pixel}$, a saturation limit of $10^\mathrm{m}$--$11^\mathrm{m}$ and a depth of ${\sim}17^\mathrm{m}$ in $V$ (\mcitealt{2017PASP..129j4502K}). Thus, GDS and \mbox{ASAS-SN} data cover different wavelength ranges at similar depths. The \mbox{ASAS-SN} ensures higher time resolution, whereas the GDS offers a higher spatial resolution. The combination of GDS and Gaia~DR3 data hold promise, since Gaia offers high-resolution space-based data, while the GDS contributes more data points per light curve and a longer time span. GDS light curves have been extensively used for the detection and characterisation of binaries and binary candidates (\mcitealt{2015A&A...582L..12H}; \mcitealt{2019MNRAS.490.5147P}; \mcitealt{2020MNRAS.493.2659S}; \mcitealt{2020A&A...644A..49M}, among others).

We describe the observation procedure, introduce the second and final data release (DR2), and provide an update on the previous variable-content study by \mcitet{2015AN....336..677K}. 

\section{Observations}\label{Sec:Obs_data}
\begin{figure*}[t]
\centerline{\includegraphics[width=1.0\textwidth]{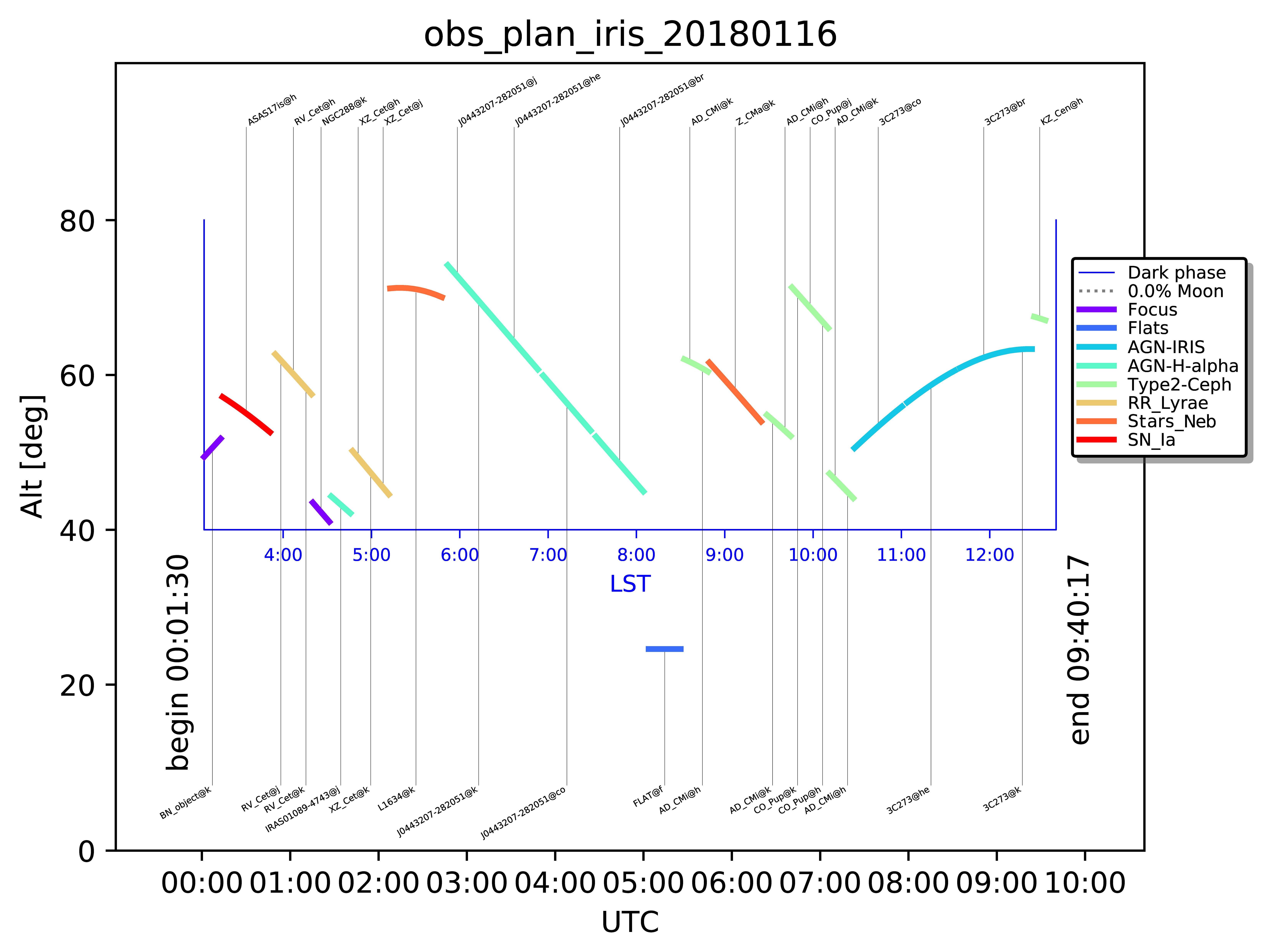}}
\caption{\enspace Observation plan created with AutoSched for IRIS, applied on 16~January~2018, displayed as elevation versus time. Further blue axes frame the time span that fulfils the required observation conditions, specifically target altitudes that lie above~$40^{\circ}$ and a limit on the Sun elevation defined for each telescope. The targets' altitude at the scheduled time is indicated by curves. The proposals corresponding to the targets are given in the legend.\label{fig:obs_plan_iris}}
\end{figure*}
\begin{figure*}[t]
\centerline{\includegraphics[width=1.0\textwidth]{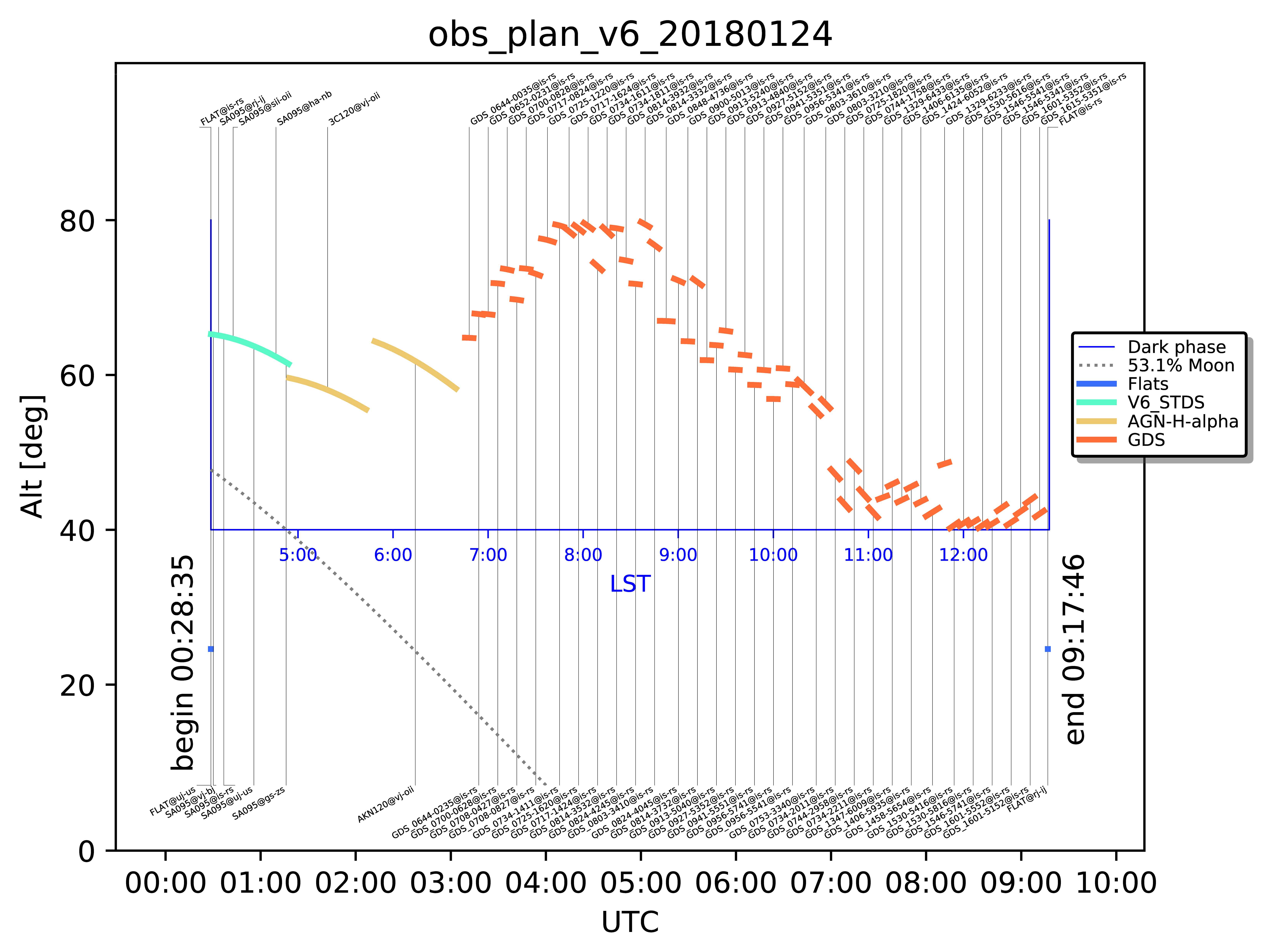}}
\caption{\enspace Observation plan created with AutoSched for RoBoTT, applied on 24~January~2018, displayed as elevation versus time, with specifications as in Figure~\ref{fig:obs_plan_iris}. \label{fig:obs_plan_v6}}
\end{figure*}
\begin{figure*}[t]
\centerline{\includegraphics[width=1.0\textwidth]{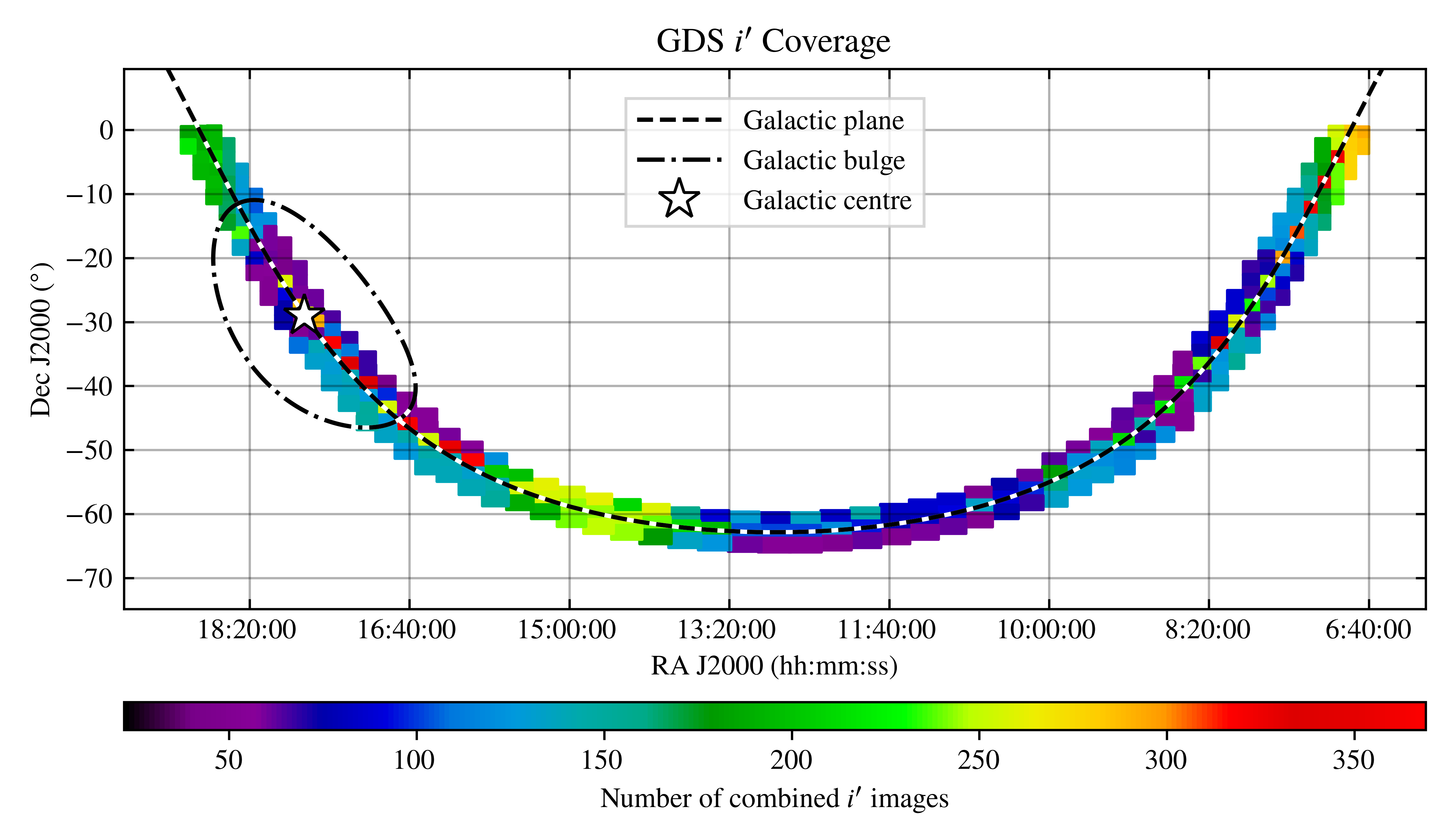}}
\caption{\enspace GDS $r'$~field coverage. \label{fig:r_coverage}}
\end{figure*}
\begin{figure*}[t]
\centerline{\includegraphics[width=1.0\textwidth]{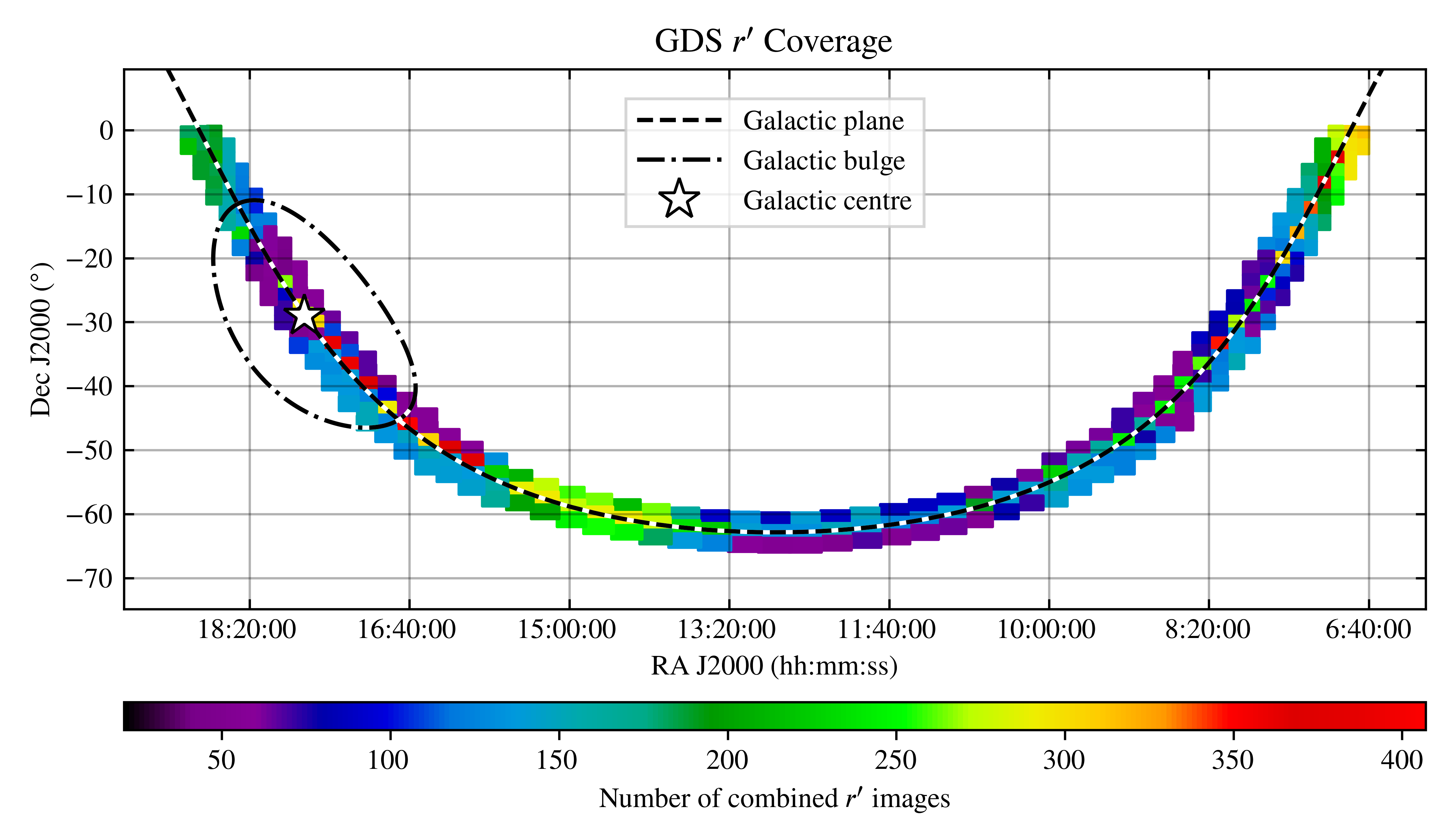}}
\caption{\enspace GDS $i'$~field coverage. \label{fig:i_coverage}}
\end{figure*}
\subsection{OCA Observation Procedure}
Observation programs at the OCM were generally long-term projects over several months or years. These observations, including GDS observations, were planned with our custom-made OCA observational scheduler and database (AutoSched), designed for the telescopes and systems that were operating since 2010, specifically the Berlin Exoplanet Search Telescope~II (BEST~II, a $25\,\mathrm{cm}$~Baker-Ritchey-Chrétien system on a German equatorial mount), the Bochum Monitoring Telescope (BMT, equipped with a $40\,\mathrm{cm}$~primary mirror and Coudé focus on an equatorial mount), formerly known as VYSOS~16 (V16), the Infrared Imaging System (IRIS, installed on a Nasmyth focus of a $80\,\mathrm{cm}$~telescope on an alt-azimuth mount), and the Robotic Bochum Twin Telescope (RoBoTT, consisting of two $15\,\mathrm{cm}$~refractor telescopes on a German equatorial mount), formerly known as the VYSOS~6 (V6), which was used for the GDS observations. Its basic principle is the creation of observation plans based on expected data quality and priorities of the target candidates, whereby the targets can be managed via a web interface. AutoSched is based on Python, the Interactive Data Language (IDL), MySQL, and PHP.

Apart from periodic types, variable observation targets require continuous monitoring with a specified minimum frequency. To meet these requirements despite missed observations (due to, e.g., bad weather) and to allow for short-term changes in cases like outbreaks, AutoSched dynamically generates observation plans for each individual night. Successful observations are recorded in databases that, in turn, are taken into account for subsequent observations. Another feature is the ability to link observations that, due to various constraints, must be conducted on the same night, ensuring they are scheduled together. Timely coordinated observations in multiple filters are necessary for methods such as reverberation mapping of active galactic nuclei, since signal and echo are captured by observations in different filters (see, e.g., \mcitealt{2018A&A...620A.137R}).

The observation procedure is summarised below.

\paragraph{\textbf{Proposal Submission}}
Proposal settings are organised in a MySQL proposal table, with a separate MySQL target table assigned to each proposal. Users receive access to their individual password-protected proposal that can be accessed from within the the OCA network. Through a PHP-based web interface, they can manage and add targets to their proposal. When creating a new target entry, the parameters right ascension, declination, filter, exposure time for a single frame $t_{\mathrm{exp}}$, and number of dithered frames to observe $n_{\mathrm{f}}$ are specified. Additional settings define the handling requirements, comprising a priority flag (translated into a priority factor $f_{\mathrm{prio}}$, with $f_{\mathrm{prio}} = 4$ for flag~A, $f_{\mathrm{prio}} = 2$ for flag~B, and $f_{\mathrm{prio}} = 1$ for flag~C), a target status (indicating whether the entry is active, pending, cancelled or done), a desired total number of observations, a repetition interval specifying when the target is due for re-observation, and a key referring to another target for synchronous observation.

Administrators can manage the proposal's status and allocated percentage of the total observation time $p_{\mathrm{prop}}$ per telescope. Besides science proposals, there are proposals that contain flat-field and standard star field observations configured by the administrators.

\paragraph{\textbf{Observation Plan Creation}}
The plan creation for each telescope starts shortly prior to the observations to allow for short-term proposal adjustments, and to ensure that previous observations are taken into account.

The start and end of the observation window are defined as the time when the Sun's upper limb touches the horizon as seen from the observation site. A framework of calibration frames, preparations, and wrap-ups is constructed for the start and end; depending on the telescope, dark, bias, dusk and dawn flats, roof opening and closing, mount control, and cooling are taken into account.

The following loop is executed for each observation time slot beginning with the first light frames that follow the calibration frames until a target's observation time would exceed the end of the observation window. A list of target candidates that meet the criteria for observation during the observation time slot is created: Targets must be part of an active proposal and marked as active by the proposer, must be visible at an altitude greater than $40^{\circ}$, and must maintain an angular distance from the Moon of at least $15^{\circ}$ to avoid overexposure. Additionally, a minimum zenith distance of $10^{\circ}$ is required for observations with IRIS because of its alt-azimuth mount, to avoid the zenith blind spot, which would otherwise require infeasibly high tracking speeds. Likewise, targets must not cross the meridian when using a telescope on a German equatorial mount to avoid a meridian flip during observations. A rank $R_{\mathrm{tar}} = p_{\mathrm{prop}} \cdot f_{\mathrm{prio}} \cdot f_{\mathrm{repeat}}$ is assigned to each target candidate; the highest ranked candidate is selected for the current time slot. If moving to the target requires a displacement of more than $5^{\circ}$ and the elapsed time since the last focusing was performed exceeds the telescope-specific threshold, refocusing is scheduled before the observation.

Through integration into the rank, the proposals' allocated percentage of observation time $p_{\mathrm{prop}}$ is not rigidly enforced each night which allows for flexibility with respect to urgency. The repetition factor $f_{\mathrm{repeat}}$ is the product of the following weighting factors: 
\begin{itemize}
    \item[--] To make sure that flat-field frames are taken each night at a favourable point of time, their $f_{\mathrm{repeat}}$ is artificially inflated to prioritise them over light frames.
    \item[--] $f_{\mathrm{repeat}}$ is increased if the elapsed time since the last observation of the target exceeds the desired repetition interval.
    \item[--] If the target's altitude is currently rising, the altitude is weighted into $f_{\mathrm{repeat}}$ with $a_\mathrm{cur} / a_\mathrm{max}$ where $a_\mathrm{max}$ is the target's maximum altitude during the observation night and $a_\mathrm{cur}$ is the current altitude. $f_{\mathrm{repeat}}$ is increased if the current altitude falls off during the target's estimated observation time~$t_{\mathrm{obs}}$.
    \item[--] If the target that is assigned to be observed synchronously with the current target is already scheduled for the current night, a time window of $3.6\,\mathrm{h}$ is allocated for scheduling its counterpart, provided it is not further limited by the repetition interval. Therefore, $f_{\mathrm{repeat}}$ is inflated to make this observation highly attractive.
\end{itemize}
The exact values of these weighting factors result from experience and testing, and are adjusted depending on the application.

The target's observation time is estimated by\linebreak$t_{\mathrm{obs}} = n_{\mathrm{f}} \cdot t_{\mathrm{exp}} + t_{\mathrm{delay}}$ with further delays $t_{\mathrm{delay}}$, which are defined as fixed durations depending on the telescope, specifically, the slew time considered when the angular distance between the current position and the target coordinates exceeds $5^{\circ}$, the guiding-initiation time, the filter-change time, the readout time, and the focusing time.

The observation plan is translated into a script following the telescope's own script design and the targets are documented in the schedule table along with their estimated observation start and end times. Examples of observation plans, including GDS observations created with Autosched for IRIS and RoBoTT, are given in Figures~\ref{fig:obs_plan_iris} and~\ref{fig:obs_plan_v6}.

\paragraph{\textbf{Observation Database and Data Storage}}
Scripts that run security backups at different physical locations are started after the completion of the observations. File names and paths, together with the observation metadata comprising target name, observation identifier, filter, and start and end times from each image's header, are recorded in the observation database. The proposed targets' settings are checked against the actual observations; if an observation block has been sufficiently completed (a minimum of 5 and at least $90\,\%$ of the required frames), the desired number of observations in the proposal table is decremented for its corresponding targets. Once the images have been stored in Bochum, they are integrated into an AFS database serving as the primary observation archive.

\subsection{GDS Observations}
The GDS has been executed from September~2010 to September~2019 with RoBoTT, which consists of two refractor telescopes, each equipped with a \mbox{$2^{\circ}42\mathrm{'} \times 2^{\circ}42\mathrm{'}$}~FoV CCD and a spatial resolution of $2.4'' / \mathrm{pixel}$. Details of RoBoTT's equipment and the OCA are presented in \mcitet{2016SPIE.9911E..2MR}.

A $6^\circ$ wide stripe along the southern Galactic disk from RA~$6\mathrm{h}40\mathrm{'}$ to RA~$19\mathrm{h}03\mathrm{'}$, comprising an area of $1323\,\mathrm{deg}^2$, is covered by the survey with 268~fields\footnote{Interactive map of the complete GDS mosaic: http://gds.astro.rub.de}; these fields were monitored regularly by simultaneous observations in $r'$ and $i'$, with corresponding centre wavelengths of $626\,\mathrm{nm}$ and $767\,\mathrm{nm}$, respectively. An observation series consists of 9~dithered images with an exposure time of $10\,\mathrm{s}$ each, yielding a magnitude range of $8^\mathrm{m} \lesssim r' \lesssim 18^\mathrm{m}$ and $7^\mathrm{m} \lesssim i' \lesssim 17^\mathrm{m}$ (see Figure~\ref{fig:GDS_mag_count} for the magnitude range of all filters). A few additional observations per field, consisting of 9~dithered images with an exposure time of $30\,\mathrm{s}$ in $UBVz'$~filters, corresponding to centre wavelengths of $365\,\mathrm{nm}$, $445\,\mathrm{nm}$, $550\,\mathrm{nm}$, and $910\,\mathrm{nm}$, allow for the determination of the source's median colours. Particular fields were additionally observed in the narrow bands $\mathrm{O}\,\mathrm{III}$, NB, $\mathrm{H}\alpha$, and $\mathrm{S}\,\mathrm{II}$ with corresponding centre wavelengths of $500.7\,\mathrm{nm}$, $645.0\,\mathrm{nm}$, $656.3\,\mathrm{nm}$, and $572.1\,\mathrm{nm}$. Regular observations of Landolt's standard star fields \mcitep{2009AJ....137.4186L} were integrated to calibrate the GDS fields' instrumental magnitudes.

The total number of combined images per field for $UBVr'i'z'$ is displayed in Figures~\ref{fig:r_coverage}, \ref{fig:i_coverage}, and~\ref{fig:UBVz_gds_coverage}. In these figures, the dimensions of the Galactic bulge are defined by a simple assumption of an ellipse with angular semi-axes of $20^\circ$ in longitude and $10^\circ$ in latitude from the Galactic centre. Observations per field in $r'$ and $i'$ by the conclusion of the survey lie between 21 and 407~combined images. All fields have at least two images in $UBVz'$. In total, 79\,210~combined images were obtained.

\section{Data Reduction, Photometry and Analysis}\label{Sec:GDSAPP}
Data reduction and astrometry for the RoBoTT data were executed using the Bochum OCA VYSOS Interactive Pipeline (BOVIP) as detailed in \mciteauthor{2012AN....333..706H} (\citeyear{2012AN....333..706H}; Paper~I). Reduced GDS data were further processed by the GDS Analysis and Photometry Pipeline (\mbox{GDS-APP}), which covers standard star analysis, photometry, calibration, construction of light curves, and variability detection among further functions, and is comprehensively described in Paper~II.

The variability criteria applied for the $r'$ and $i'$~light curves, i.e., standard deviation, amplitude, and Stetson's \mbox{$J$-Index} \mcitep{1996PASP..108..851S}, were complemented by magnitude dependencies since Paper~II. While the lower limit for the number of data point to check for variability generally was~10, this limit was not applied for $i'$ in the field GDS\_1808-2411. From 29~combined images for this filter-field combination, only 9~images turned out to be usable for the creation of light curves. Owing to an irreparable failure of one of RoBoTT's CCDs, the resumption of observations is not possible. To avoid compromising the completeness of variability mapping across the GDS mosaic, reduced accuracy of variability detection is considered acceptable in this case.

\section{GDS Catalogues}
\subsection{Image Catalogue}\label{Sec:GDS_imgs}
Reduced, uncropped images\footnote{https://dc.g-vo.org/BGDS} of the GDS are publicly available via the German Astrophysical Virtual Observatory\footnote{https://www.g-vo.org} (GAVO). We have supplemented the image catalogue with reduced observations taken between 2017 and 2019, and updated a portion of the original image data release. Besides the GDS fields, standard star field observations used for the calibration of all RoBoTT surveys including the GDS were added. The complete image data release covers the GDS field and standard star field images throughout the entire survey time span. It contains images with sufficient data quality for astrometry including those images that did not pass quality control for further analysis with the \mbox{GDS-APP}.

\subsection{Light Curve and Photometry Catalogues}\label{Sec:LC_Phot_cat}
Light curves are constructed for sources with at least 5 available data points; otherwise, only the sources' median magnitudes together with the corresponding observation dates are specified. $r'$ and $i'$~light curves are available for all 268~fields. For most of the fields, the quality of the $UBVz'$~images is sufficient for photometry. Initially, light curves were only available in $r'$ and $i'$. After further observations in $UBVz'$ and additional observations in the narrow bands, we have obtained light curves for those filters in specific fields. Although light curves are available for these additional filters, only the $r'$ and $i'$ light curves were analysed for variability in accordance with the variability criteria mentioned in Section~\ref{Sec:GDSAPP} because they were observed most frequently.

Light curves and photometry are stored in separate tables for each filter; there is no association between sources across filters. A portion of the sources has multiple light curves per filter: The central coordinates of the fields are $2.0^{\circ}$ apart. Since the images are cropped to $2.2^{\circ} \times 2.2^{\circ}$, sources in the overlapping coordinate range will be observed in multiple fields. As described in \mbox{Papers~I and II}, environmental influences that lead to night-to-night fluctuations in the light curves are corrected separately for each field. In principle, this approach leads to offsets between fields, which usually lie below $\pm\,0.05^\mathrm{m}$. Since the main purpose of the GDS is the identification and classification of variables and this method for correction of night-to-night fluctuations has been shown to be reliable for restoring the shape of the light curves, these errors in the absolute magnitude are acceptable and are taken into account by the uncertainties shown in the catalogues. However, offsets can be noticeable when comparing light curves from different fields and may hamper period determination depending on the algorithm used. Therefore, data from different fields were not combined by default and sources in overlapping zones have separate light curves from each contributing field.

For sources with available light curves, the photometry catalogue contains median magnitudes derived from the corresponding light curves. Similarly to the the light curves, median magnitudes are not averaged over multiple fields.

For almost 21.7~million sources, magnitudes are measured in at least one filter. Light curve and photometry databases of the complete survey, including usage examples, are available from the VO (\mcitealt{blexetal2024a}; \citeyear{blexetal2024b}) as well as in the form of browser services\footnote{https://dc.g-vo.org/BGDS2-lc, https://dc.g-vo.org/BGDS2-med}. An overview page, listing all services and tables for the complete data release\footnote{http://dc.g-vo.org/browse/bgds/l2\label{footnote:dr2_overview}} alongside access options, is also provided. It should be noted that, owing to the inclusion of additional standard star field observations, magnitudes differ slightly from those in the first release; the first release of the light curve and photometry data remains available\footnote{http://dc.g-vo.org/browse/bgds/l}. 

\section{Content Overview and Properties}
\begin{figure*}[htbp]
\centerline{\includegraphics[width=1.0\textwidth]{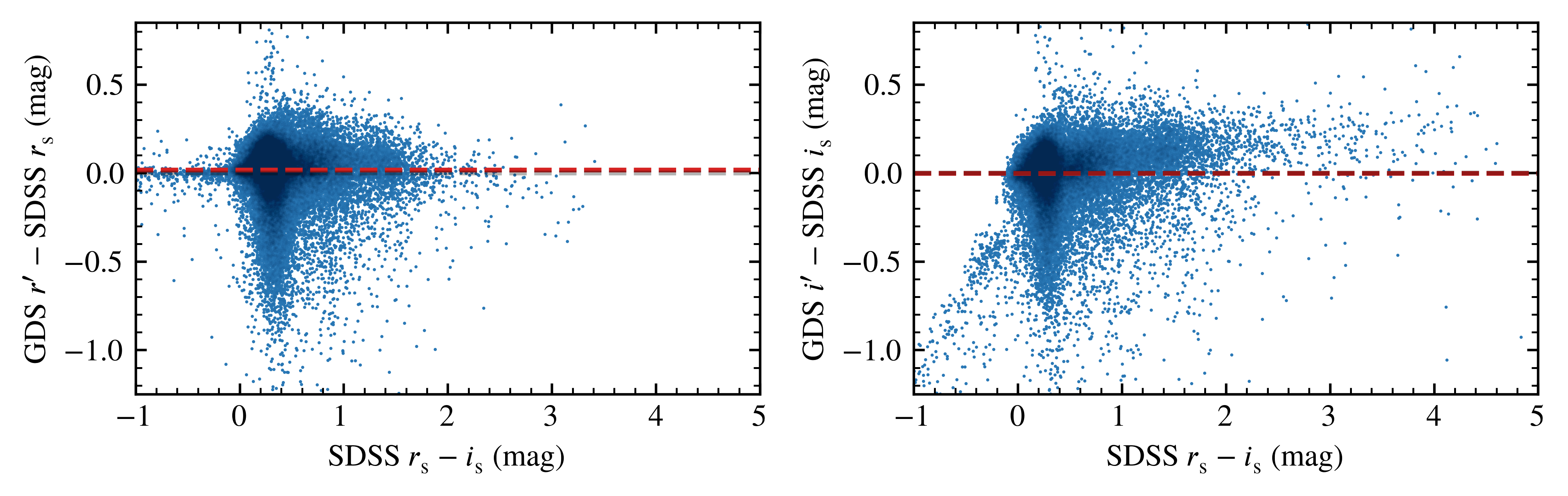}}
\caption{\enspace Offset of GDS~$r'$ from SDSS~$r_\mathrm{s}$ (left) and GDS~$i'$ from SDSS~$i_\mathrm{s}$ (right) as a function of SDSS~$r_\mathrm{s}-i_\mathrm{s}$ colour for the physically plausible $r_\mathrm{s}-i_\mathrm{s}$~colour range of $-1^\mathrm{m}$ to $5^\mathrm{m}$ for 69\,292~sources in $r'$ and 45\,046 in $i'$ with clean photometry flags in the SDSS; offsets obtained by least squares fitting of the data in the respective plotted range are indicated by red dashed lines. \label{fig_sdss_ri}}
\end{figure*}
\begin{figure}[htbp]
	\centerline{\includegraphics[width=1.0\columnwidth]{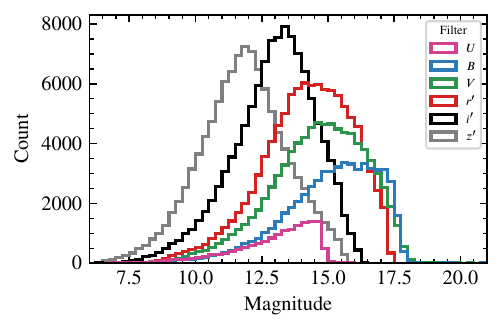}}
	\caption{\enspace Magnitude distribution of variable sources. \label{fig:GDS_mag_count}}
\end{figure}
\begin{figure}[htbp]
	\centerline{\includegraphics[width=1.0\columnwidth]{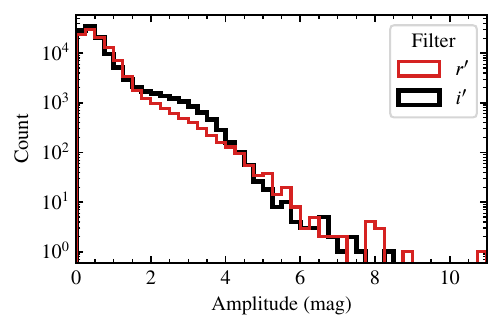}}
	\caption{\enspace $r'i'$~amplitude distributions of variable sources.\label{fig:GDS_amp_count}}
\end{figure}
\begin{figure}[htbp]
	\centerline{\includegraphics[width=1.0\columnwidth]{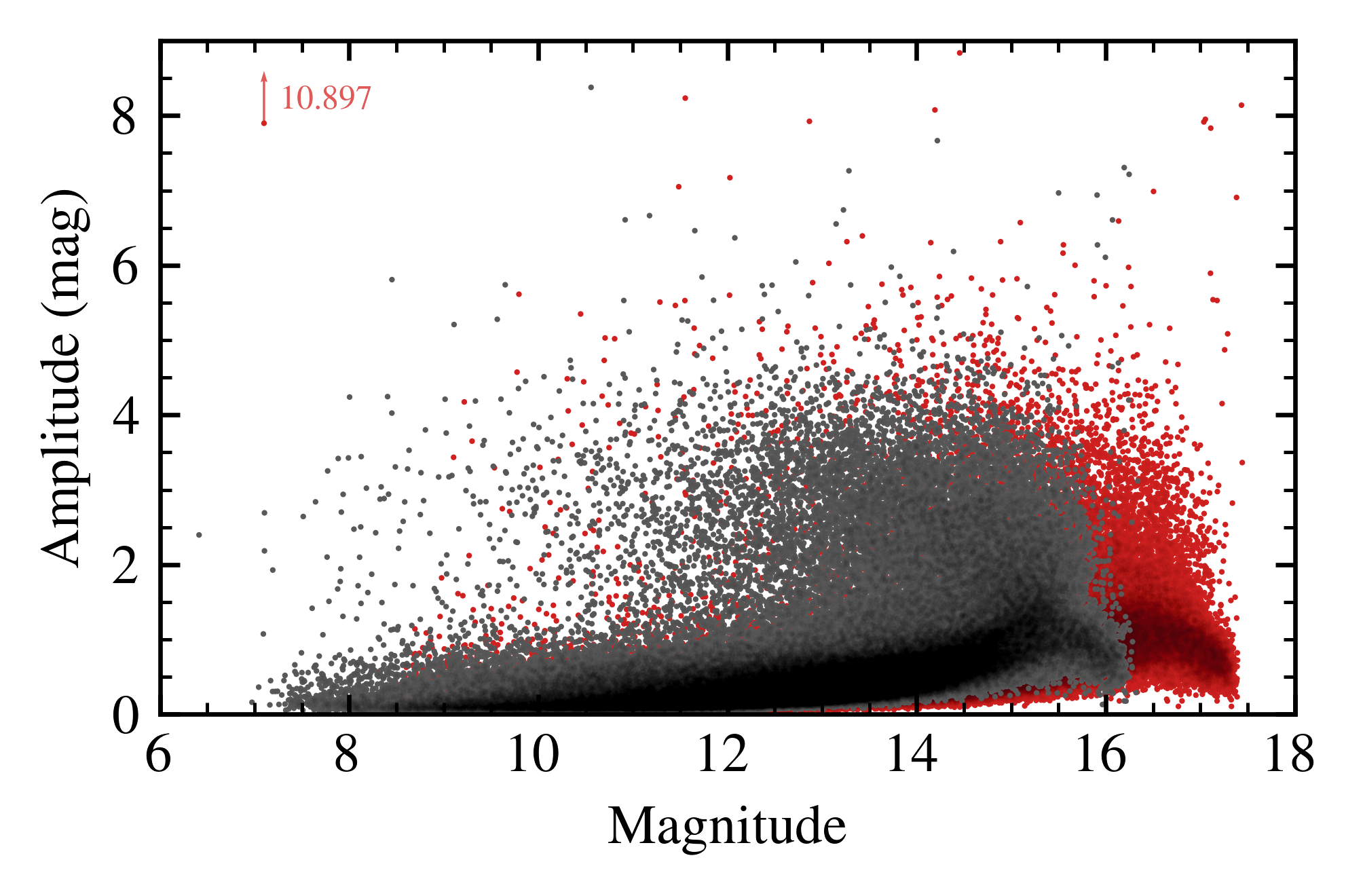}}
	\caption{\enspace Brightness dependency of amplitudes in $r'$ (red) and $i'$ (grey) of variable sources.\label{fig:GDS_mag_amp}}
\end{figure}
\begin{figure*}[htbp]
\centerline{\includegraphics[width=1.0\textwidth]{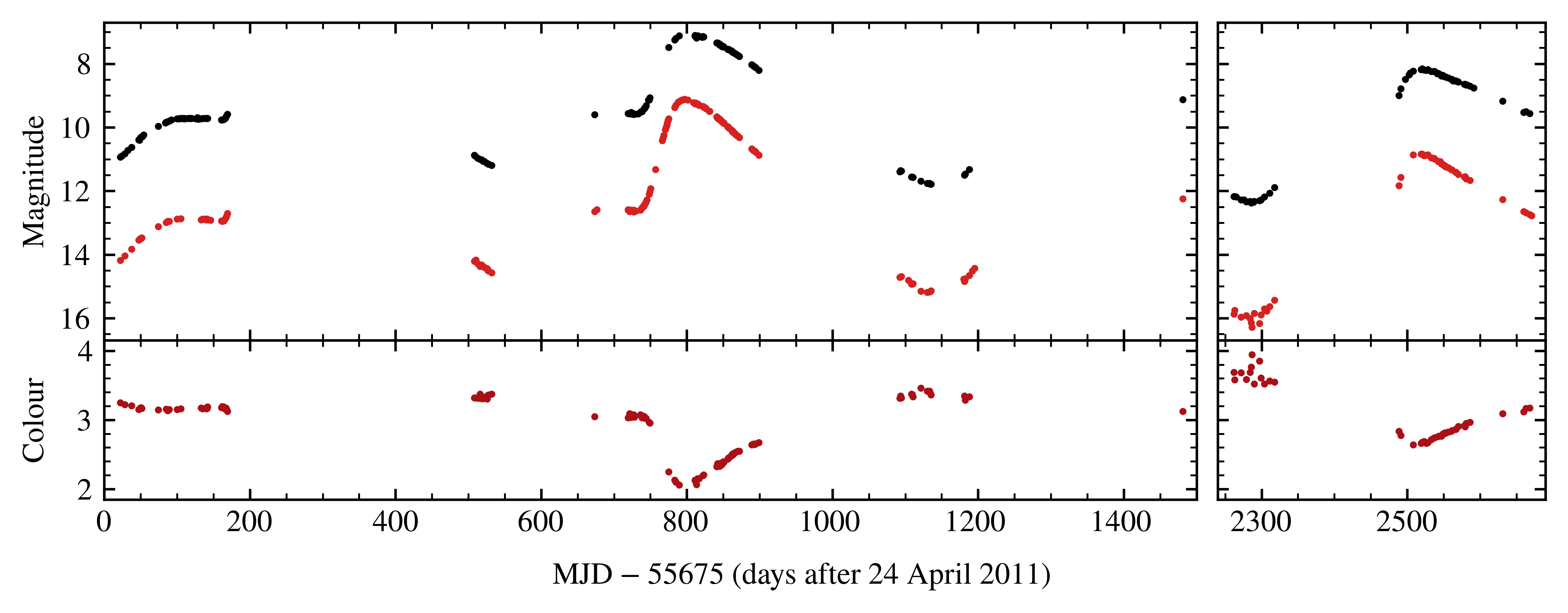}}
\caption{\enspace $r'$ (red) and $i'$ (black) light curves and $r' - i'$~colour curve of V352~Nor.\label{fig:LCV345Nor}}
\end{figure*}
\begin{figure}[htbp]
	\centerline{\includegraphics[width=1.0\columnwidth]{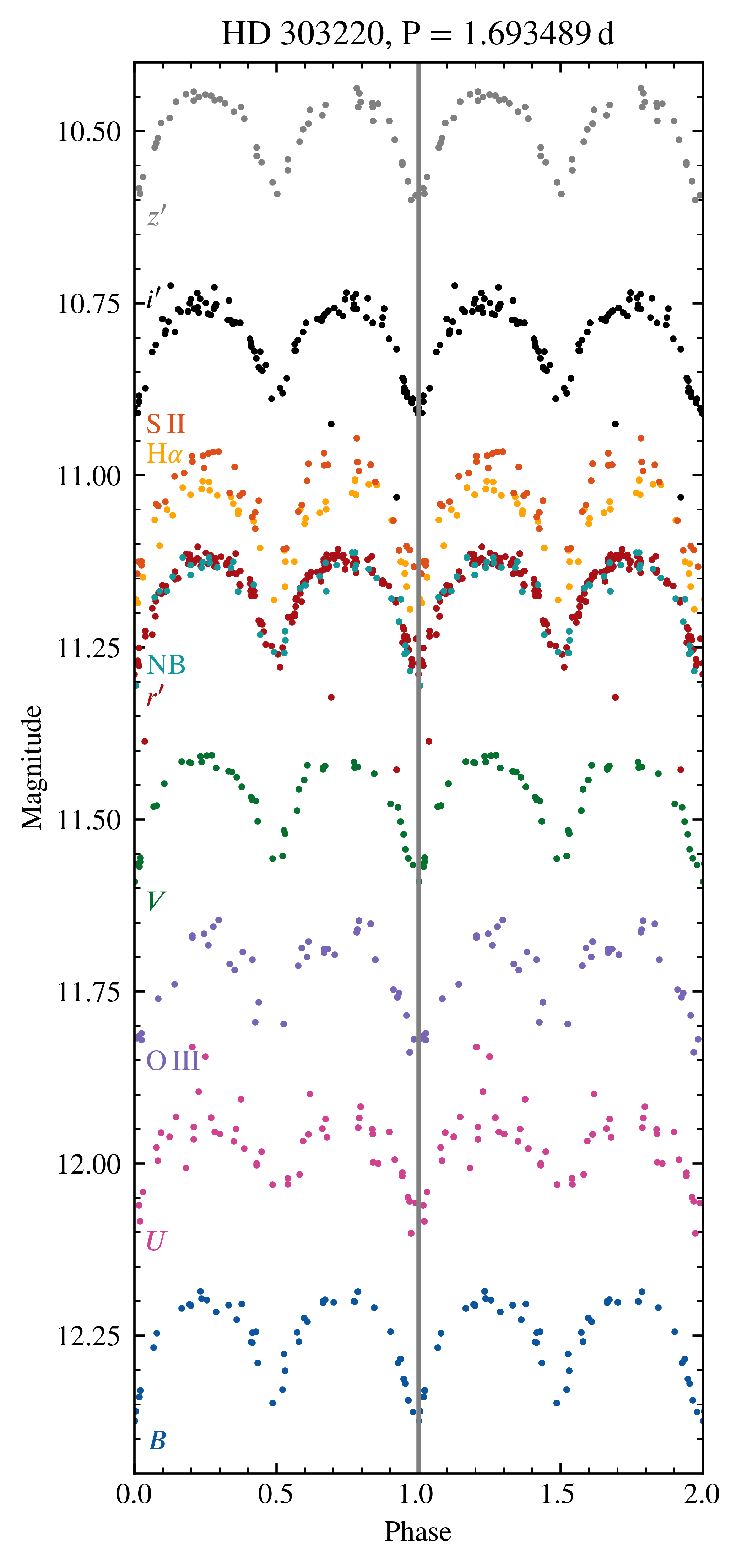}}
	\caption{\enspace Folded light curves in $UBVr'i'z'$ and the narrow bands $\mathrm{O}\,\mathrm{III}$, NB, $\mathrm{H}\alpha$, $\mathrm{S}\,\mathrm{II}$ of HD~303220.\label{Fig:multi-lcs}}
\end{figure}
The photometry and spatial distribution of the GDS sources are compared with Sloan Digital Sky Survey (SDSS) photometry and Infrared Astronomical Satellite (IRAS) images, respectively, in Sections~\ref{Sec:Phot}--\ref{Sec:SpatDist}. In Section~\ref{Sec:Var_cont}, we provide a statistical overview of the variable content and show representative and particularly prominent light curve examples in Section~\ref{Sec:LC_ex}.

Because sources in different filter-field combinations are not cross-matched within the light curve and photometry catalogues, they are therefore subsequently supplemented by cross-filter matching information and spectral classes from the multi-wavelength catalogue\footnote{https://dc.g-vo.org/BGDS2-matched}, which will be introduced in full detail in the next part of this series. In the multi-wavelength catalogue, sources of every filter-field combination are cross-matched with Gaia~DR3 (\mcitealt{2016A&A...595A...1G}; \citeyear{2023A&A...674A...1G}; \mcitealt{2023A&A...674A..32B}) sources. Spectral classes are estimated from GDS $UBV$ photometry and compiled from SIMBAD and the International Variable Star Index\footnote{https://www.aavso.org/vsx/} (VSX; \mcitealt{2006SASS...25...47W}).

\subsection{Photometry}\label{Sec:Phot}
The small overlap between the GDS and SDSS fields enables documenting the colour-dependent offset between the GDS photometry and the stable photometry of the SDSS for all sources with matching magnitudes in the respective filters. The comparison of GDS and SDSS magnitudes displayed in Figure~\ref{fig_sdss_ri} is based on a total of 358\,458 matches, independent of available filter photometry and its quality flags, within a $3''$~radius against the 16th data release of the SDSS (DR16; \mcitealt{2020ApJS..249....3A}); VizieR catalogue \texttt{V/154}). A minor offset is inevitable, since the detectors and the $r'$ and $i'$~filters used in the GDS differ from those of the SDSS photometric system \mcitep{1996AJ....111.1748F}.

\subsection{Spatial Distribution}\label{Sec:SpatDist}
Stars that have been observed in at least one filter and specifically the subset of variable stars are mapped in Galactic coordinates in Figure~\ref{fig:gds_iras}. These are complemented by the 60~Micron Band data from the IRAS Sky Survey Atlas (ISSA) Galactic Plane Mosaics\footnote{https://lambda.gsfc.nasa.gov/product/iras/iras\_issam\_get.html} to trace gas density and cool dust. Consistent with expectations, regions that show high emission in the 60~Micron Band are sparsely populated with stars detectable by the GDS because of reddening. Many of the observed stars lie outside the Galactic plane and are supplemented with a small number of foreground stars.

\subsection{Variable Star Content}\label{Sec:Var_cont}
The additional observations since 2015 have led to an increase from the previous intermediate result of 64\,151~variable sources that met the variability criteria in Paper~II to a total of 113\,449~variable sources. This number refers to a lower bound because multiple sources located within a radius of $3''$, which could not be definitely distinguished owing to the limited resolution of the RoBoTT images, were counted only once. To maintain reliability, only the highest-confidence classifications are included; indeed, additional variables can be found in the complete light curve catalogue (e.g., \mbox{TYC\_4813-2867-1} in Figure~\ref{fig:LcsUnknownVar}). 20\,798 of the sources classified as variables are not listed in either the VSX or the cross-match catalogue of Gaia sources and literature variables compiled by \mcitet{2023A&A...674A..22G}. The variable sources that were included in the GDS version presented in Paper~II have already been recorded in the VSX. Accounting for the newly identified variable sources from Paper~II yields 77\,592~new variables that were discovered over the entire survey period.

As a result of re-evaluating all detected sources using the variability criteria, not only does an increased number of sources meet the threshold for variability classification, but some previously identified variables can not be confirmed when considering the expanded dataset. Accordingly, cross-matching the current set of variable stars against the earlier catalogues from Paper~I (VizieR catalogue \texttt{J/AN/333/706}) and Paper~II (VizieR catalogue \texttt{J/AN/336/590}) using a $3''$~matching radius yields 528 and 17\,387~discarded sources, respectively.

Among the variable sources, photometry in $UBVr'i'z'$ is available for approximately $13.3\,\%$, $51.0\,\%$, $74.9\,\%$, $93.0\,\%$, $99.1\,\%$, and $94.5\,\%$ of the sources, respectively; the corresponding magnitude distributions are shown in Figure~\ref{fig:GDS_mag_count}. Notable differences in the magnitude cut-offs between filters in Figure~\ref{fig:GDS_mag_count} may originate from the atmospheric transmission that affects the effective filter transmission at the long-wavelength flank of the filters. Particularly, the $z'$~filter's wavelength range is dominated by OH and $\mathrm{H}_{2}\mathrm{O}$ absorption, which could explain the more extended flank on the fainter side of the $z'$ distribution. Another factor contributing to the earlier magnitude cut-off in $U$ is that the CCDs are less sensitive at these wavelengths.

Magnitude and amplitude distributions in $r'$ and $i'$ of the survey's variable star content are plotted in Figures~\ref{fig:GDS_amp_count}--\ref{fig:GDS_mag_amp}. The $r'$ and $i'$~filters are used as illustrative examples because their median magnitudes are less sensitive to variations owing to the larger number of measurements, and they enable a comparison with the previous version (see Figures~8 and~9 in Paper~II, and Figure~4 in \mcitealt{2015AN....336..677K}) for which $UBVz'$~information was not fully available.

Due to the offsets between light curves of different fields as mentioned in Section~\ref{Sec:LC_Phot_cat}, deriving amplitudes from combined light curves constructed from observations of multiple fields is discouraged. In cases where light curves from multiple fields are available for a given source, amplitudes were derived from the light curves yielding the highest values for use in Figure~\ref{fig:GDS_amp_count} and all subsequent figures displaying amplitudes.

The incorporation of magnitude dependencies into the variability criteria partially mitigates for the paucity of detected low-amplitude variables at faint magnitudes. This section is now more densely populated (see Figure~\ref{fig:GDS_mag_amp}) as a compromise between compensation for poorer detection conditions and misclassifications due to artificial errors and other causes. The measured $r'$ and $i'$~amplitudes lie between $0.01^\mathrm{m}$ and $10.90^\mathrm{m}$; the highest observed amplitude was recorded for V906 Car, a nova discovered by \mcitet{2018ATel11454....1S}.

\subsection{Light Curves of Selected Sources}\label{Sec:LC_ex}
To demonstrate the range in data quality, a selection of folded light curves for periodic variables across the observed magnitude range is presented in Figures~\ref{fig:LcsKnownPer}--\ref{fig:LcsUnknownVar}. Periods were determined using the Lafler-Kinman algorithm \mcitep{1965ApJS...11..216L}. Figure~\ref{fig:LcsKnownPer} shows the best possible case, where the maximum potential is realised due to a large number of data points and large amplitude. The light curves in Figure~\ref{fig:LcsKnownVar1} are not equally well defined; however, we were still able to extract periods that were previously either unknown or incorrectly determined, as recorded in the VSX and VizieR \mcitep{2000A&AS..143...23O} for known variables. Examples of light curves for variable stars previously unknown to the VSX and VizieR are shown in Figure~\ref{fig:LcsUnknownVar}. The variability and periods were checked for the coordinates of the GDS sources within a radius of $10''$. In the case of the Stellar variability in Gaia~DR3 catalogue (\mcitealt{2023A&A...677A.137M}; VizieR catalogue \texttt{J/A+A/677/A137}), only the Gaia~DR3 photometric variability flag, but not the variability flags for particular filters, was considered. Presumably, the subset from which the examples in Figure~\ref{fig:LcsKnownVar1} are drawn is most interesting for further investigations, because the light curve quality is in many cases not only sufficient for the determination of periods, but also allows for new variability type classifications.

The particularly interesting example of V352~Nor is shown in Figure~\ref{fig:LCV345Nor}, demonstrating the worth of continued long-term observations. This star was classified as a long-period variable and Mira variable candidate \mcitep{1996IBVS.4403....1L} with a period of $576\,\mathrm{d}$ \mcitep{2012A&A...548A..79A} as well as a symbiotic variable \mcitep{2000A&AS..146..407B}. This source exhibits continuous variability during our observations, allowing us to detect changes in its colour, precisely tracked by the $r'$ and $i'$~light curves. The tight scatter seen in Figures~\ref{fig:LcsKnownPer}--\ref{fig:LcsUnknownVar} and~\ref{fig:LCV345Nor} is a consequence of rigorous quality control; together with the large number of observations in $r'$ and $i'$, this yields reliable amplitude measurements for the variable sources in both filters.

As mentioned previously, particular GDS fields were observed repeatedly in multiple filters, enabling the construction of light curves in each filter, an example of which is the phase plot in Figure~\ref{Fig:multi-lcs} displaying the light curves of the eclipsing binary HD~303220.

\section{Summary and Outlook}
The GDS has concluded after an observation span of more than nine years, yielding light curves for 113\,449~variable sources, of which more than two thirds were newly identified from the GDS light curves. The given overview implies that the GDS light curves are spread across various variability types and are suitable for classifying and investigating them independently or in combination with \mbox{ASAS-SN} light curves (\mcitealt{2017PASP..129j4502K}; \mcitealt{2018MNRAS.477.3145J}), for instance. To this end, all GDS data sets were made publicly available to the scientific community via the GAVO\footref{footnote:dr2_overview}. GDS data catalogues or light curve data packages can also be received directly from the authors. These catalogues comprise 79\,210~images across 268~fields, photometry in one to ten filters for almost 21.7~million sources, and light curves for the majority of these sources.

The current data structure, which combines light curves and photometry based on the number of images available for specific filters within each field, is intrinsically complex. Despite the higher resolution achieved by resampling the combined images, cross-matching data from different filters within the survey or against external databases remains cumbersome in crowded regions. In addition, the previously noted field overlap may result in up to four light curves or magnitudes per filter for a given source. This issue will be addressed in the next part of this series, which introduces the multi-wavelength catalogue intended to provide a more useful, user-friendly resource.

In the near future, data products from several surveys that comprise the Milky Way's stellar content are expected. The Data Release 1 of the eROSITA All-Sky Survey (eRASS~DR1; \mcitealt{2024A&A...682A..34M}) that includes almost $930\,000$~entries for the $0.2$--$2.3\,\mathrm{keV}$~energy range was published at the beginning of 2024; the following release is slated for 2026. Upcoming surveys that cover different wavelength ranges will synergise with the time series data provided by the GDS.

Concerning high-mass stars, the Javalambre Photometric Local Universe Survey (J-PLUS; \mcitealt{2019A&A...622A.176C}), which is being conducted with the \mbox{JAST/T80} telescope at the Observatorio Astrofísico de Javalambre, uses a specific system of twelve filters to retrieve spectral energy distributions (SEDs). \mbox{J-PLUS's} filter set is designed to capture stellar-type information, such as the Balmer break region, which makes these SEDs suitable for stellar classifications and enables an effective search for O~stars, in particular. The current third data release\footnote{https://www.j-plus.es/datareleases/data\_release\_dr3} covers $3192\,\mathrm{deg}^2$ of the northern hemisphere and more than 47~million objects. This serves as an example of the valuable contributions by smaller telescopes.

With first light already achieved, the Vera C. Rubin Observatory is preparing to begin its 10-year Legacy Survey of Space and Time (LSST; \mcitealt{2019ApJ...873..111I}), which targets the southern sky. The LSST aims for a depth of $r > 27^\mathrm{m}$ as a requirement to investigate the main-sequence stars up to the edge of the halo \mcitep{2019ApJ...873..111I}. Conversely, the saturation limit for the LSST is estimated to lie at $r {\sim} 16^\mathrm{m}$ \mcitep{2019ApJ...873..111I}.

Among the ongoing an upcoming projects within the GDS's wavelength and magnitude range are the ZTF, the Asteroid Terrestrial-impact Last Alert System (ATLAS; \mcitealt{2018PASP..130f4505T}; \mcitealt{2020PASP..132h5002S}), the BlackGEM array \citep{2024PASP..136k5003G}, and the La Silla Schmidt Southern Survey (LS4; \mcitealt{2025PASP..137i4204M}). While the ZTF observes the northern sky, ATLAS (with its telescopes in South Africa and Chile commissioned in December 2021 and January 2022), the BlackGEM array (in operation since May 2023), and the upcoming LS4 target the southern sky and began or will begin after the GDS concluded in September 2019. Therefore, the GDS will remain a relevant legacy resource for variability data in the visual to short near-infrared wavelength range and for extending the time baselines of later surveys in the southern sky.

Besides the GDS, observations for several other projects were conducted with RoBoTT. The majority of these observations were single‑pointing rather than mosaics; their target fields focused on specific objects, mainly active galactic nuclei and Milky Way stars. Apart from a few exceptions, only these central sources have been analysed so far, while all other sources in the \mbox{$2^{\circ}42\mathrm{'} \times 2^{\circ}42\mathrm{'}$}~images remain unused. Therefore, we are preparing these additional RoBoTT images for a unified data release that will be provided by the GAVO.

\section*{Acknowledgements}
Standards-compliant publication and browser-based retrieval of the data collections described here is provided by the GAVO data centre, operated by e-inf-astro (BMBF Förderkennzeichen 05A23VH4).

This work was supported by the Nordrhein–Westfälische Akademie der Wissenschaften und der Künste, funded by the Federal State Nordrhein-Westfalen and the Federal Republic of Germany.

This research has made use of the SIMBAD database, operated at CDS, Strasbourg, France.
This research has made use of the VizieR catalogue access tool, CDS, Strasbourg, France (DOI : 10.26093/cds/vizier). The original description of the VizieR service was published in 2000, A{\&}AS 143, 23. This research made use of the cross-match service provided by CDS, Strasbourg. This research has made use of the International Variable Star Index (VSX) database, operated at AAVSO, Cambridge, Massachusetts, USA.

This work has made use of data from the European Space Agency (ESA) mission {\it Gaia} (\url{https://www.cosmos.esa.int/gaia}), processed by the {\it Gaia} Data Processing and Analysis Consortium (DPAC, \url{https://www.cosmos.esa.int/web/gaia/dpac/consortium}). Funding for the DPAC has been provided by national institutions, in particular the institutions participating in the {\it Gaia} Multilateral Agreement. The IRAS ISSA Galactic Plane Mosaics were produced by Drs.~Gaylin~Laughlin and Rick~Ebert as part of a NASA Astrophysics Data Program research effort led by Dr.~William~Waller.

We thank Martin~Haas for providing us with the most recent version and instructions for using the BOVIP. The Tool for OPerations on Catalogues And Tables\footnote{http://www.starlink.ac.uk/topcat/} (\mbox{TOPCAT}; \mcitealt{2005ASPC..347...29T}) and its command-line counterpart, the Starlink Tables Infrastructure Library Tool Set\footnote{http://www.starlink.ac.uk/stilts/} (STILTS; \mcitealt{2006ASPC..351..666T}), were used for cross-matching. This work made extensive use of Python together with the packages \texttt{NumPy} \mcitep{harris2020array}, \texttt{pandas} (\mcitealt{reback2020pandas}; \mcitealt{mckinney-proc-scipy-2010}), \texttt{matplotlib} \mcitep{Hunter:2007}, \texttt{seaborn} \mcitep{Waskom2021}, and \texttt{PyEphem}\footnote{https://rhodesmill.org/pyephem/} \mcitep{2011ascl.soft12014R}. This work made use of Astropy:\footnote{http://www.astropy.org} a community-developed core Python package and an ecosystem of tools and resources for astronomy (\mcitealt{2013A&A...558A..33A}; \citeyear{2018AJ....156..123A}; \citeyear{2022ApJ...935..167A}).

We thank the anonymous reviewer for constructive comments.

\subsection*{Conflicts of Interest}
The authors declare no conflicts of interest.

\subsection*{Data Availability Statement}
The GDS data catalogues are openly available in the GAVO:
\begin{flushleft}
\begin{itemize}
    \item[--] Bochum Galactic Disk Survey images: \\
https://dc.g-vo.org/BGDS
    \item[--] BGDS~DR2 light curves: \\
https://doi.org/10.21938/dChYyzuCGA00rsfzfQ8v:Q
    \item[--] BGDS~DR2 Median Photometry Cone Search: \\
https://doi.org/10.21938/vjJw6sq0gmS5psH9uM5\_NA
    \item[--] BGDS~DR2 Matched Photometry Cone Search: \\
https://doi.org/10.21938/whiHSDPrL:m8cI6AWZ:CFg
\end{itemize}
\end{flushleft}

\bibliography{GDS_III}
\appendix
\section{Coverage\label{app1}}
\section{Light Curves\label{app2}}
\renewcommand{\thefigure}{A\arabic{figure}}
\begin{sidewaysfigure*}
\begin{center}
\centerline{\includegraphics[width=1.0\textheight]{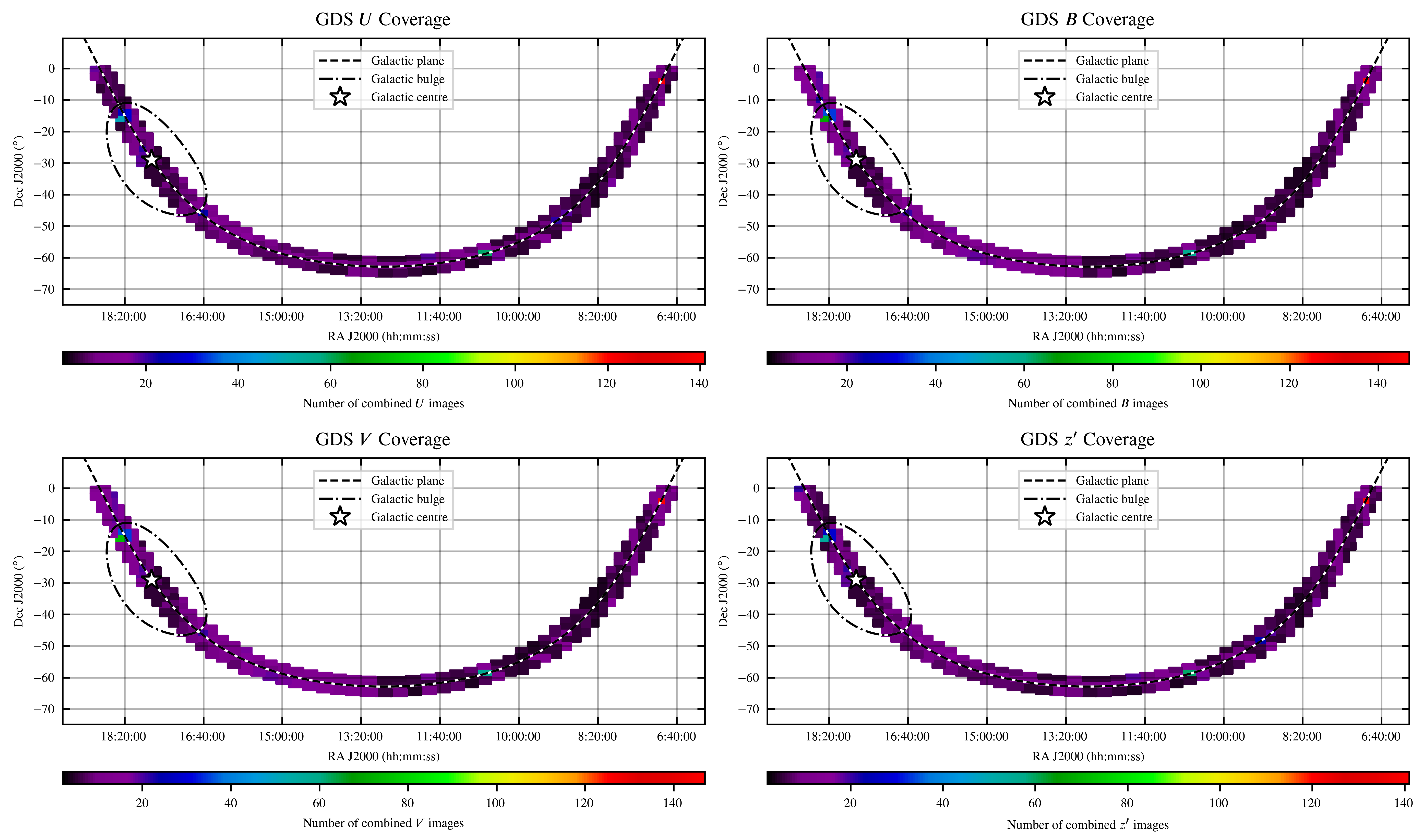}}
\caption{\enspace GDS $UBVz'$~field coverage. \label{fig:UBVz_gds_coverage}}
\end{center}
\end{sidewaysfigure*}
\begin{sidewaysfigure*}
\begin{center}
\centerline{\includegraphics[width=1.0\textheight]{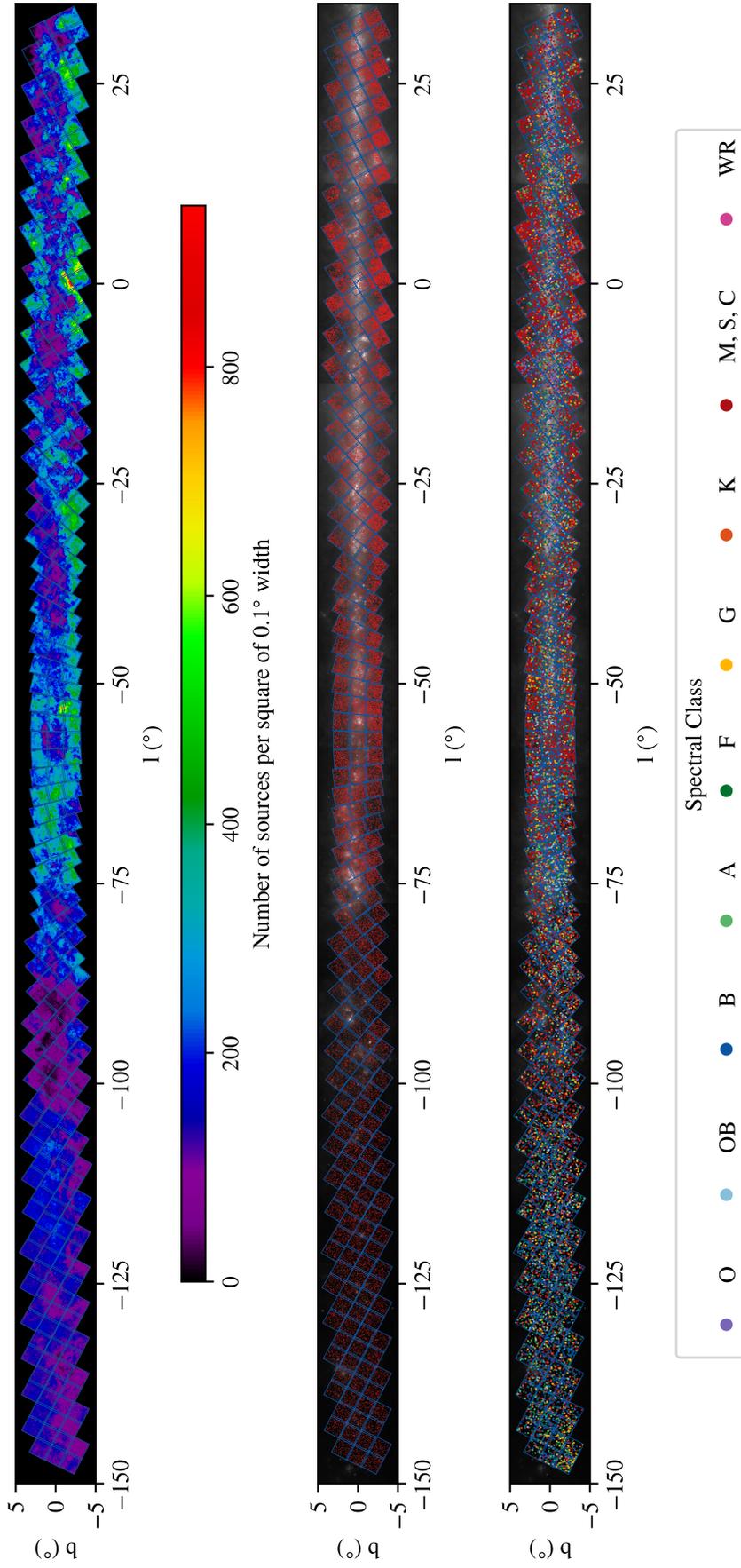}}
\caption{\enspace Galactic coordinates of GDS sources, overlaid with the cropped GDS field borders (blue). In the density map (top) all GDS sources from the multi-wavelength catalogue included, whereas the source distributions incorporate the variable GDS sources (middle) and variable stars with available spectral information (bottom). The source distributions are underlaid with the IRAS 60~Micron Band, which is logarithmically scaled to emphasise the dust structure for illustrative purposes only. \label{fig:gds_iras}}
\end{center}
\end{sidewaysfigure*}
\renewcommand{\thefigure}{B\arabic{figure}}
\begin{figure*}[p]
\centerline{\includegraphics[width=1.0\textwidth]{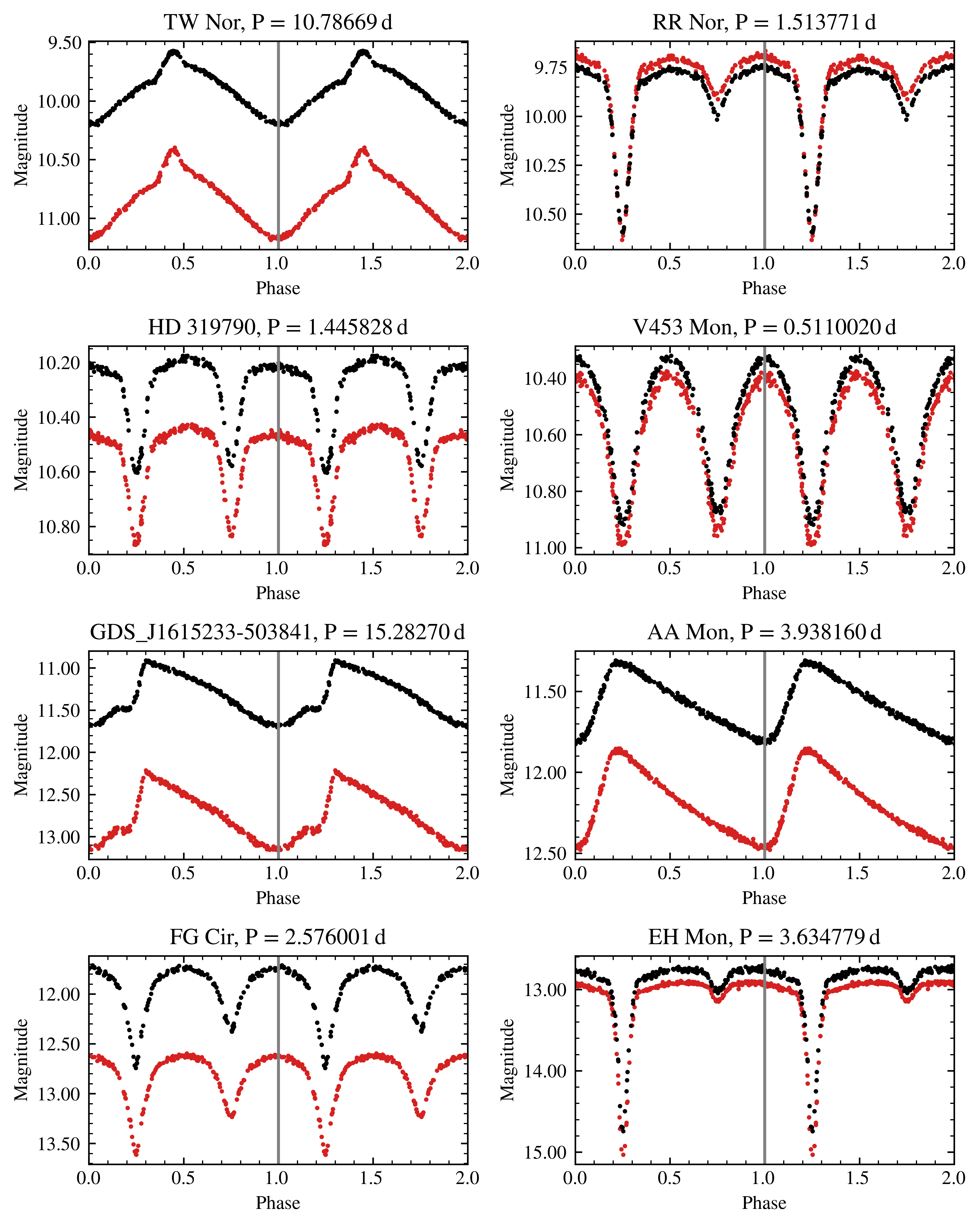}}
\caption{\enspace Folded $r'$ (red) and $i'$ (black) light curves for a selection of stars whose periods were previously known.\label{fig:LcsKnownPer}}
\end{figure*}
\begin{figure*}[p]
\centerline{\includegraphics[width=1.0\textwidth]{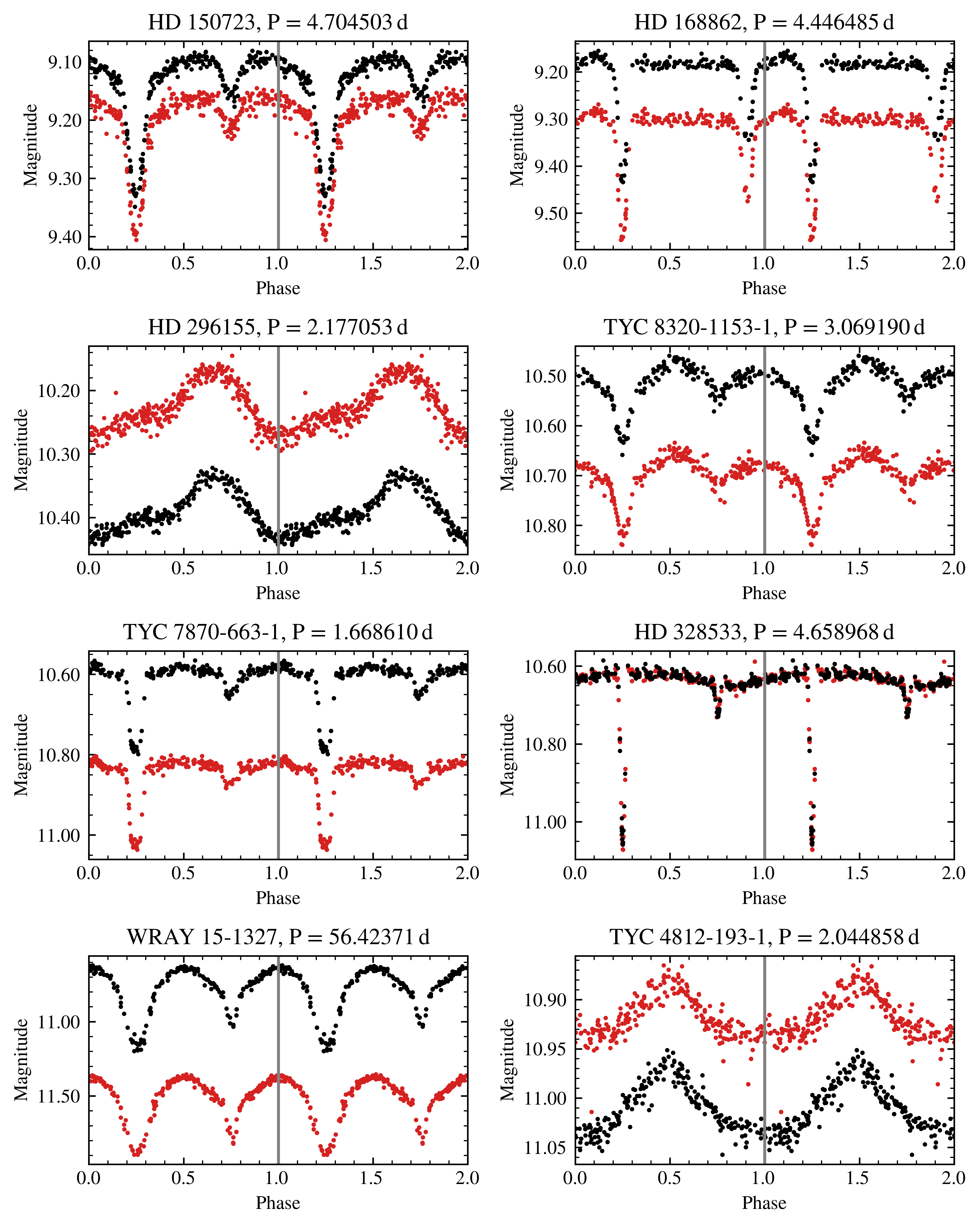}}
\caption{\enspace Folded $r'$ (red) and $i'$ (black) light curves for a selection of stars whose periods were previously unknown.\label{fig:LcsKnownVar1}}
\end{figure*}
\begin{figure*}[p]
\centerline{\includegraphics[width=1.0\textwidth]{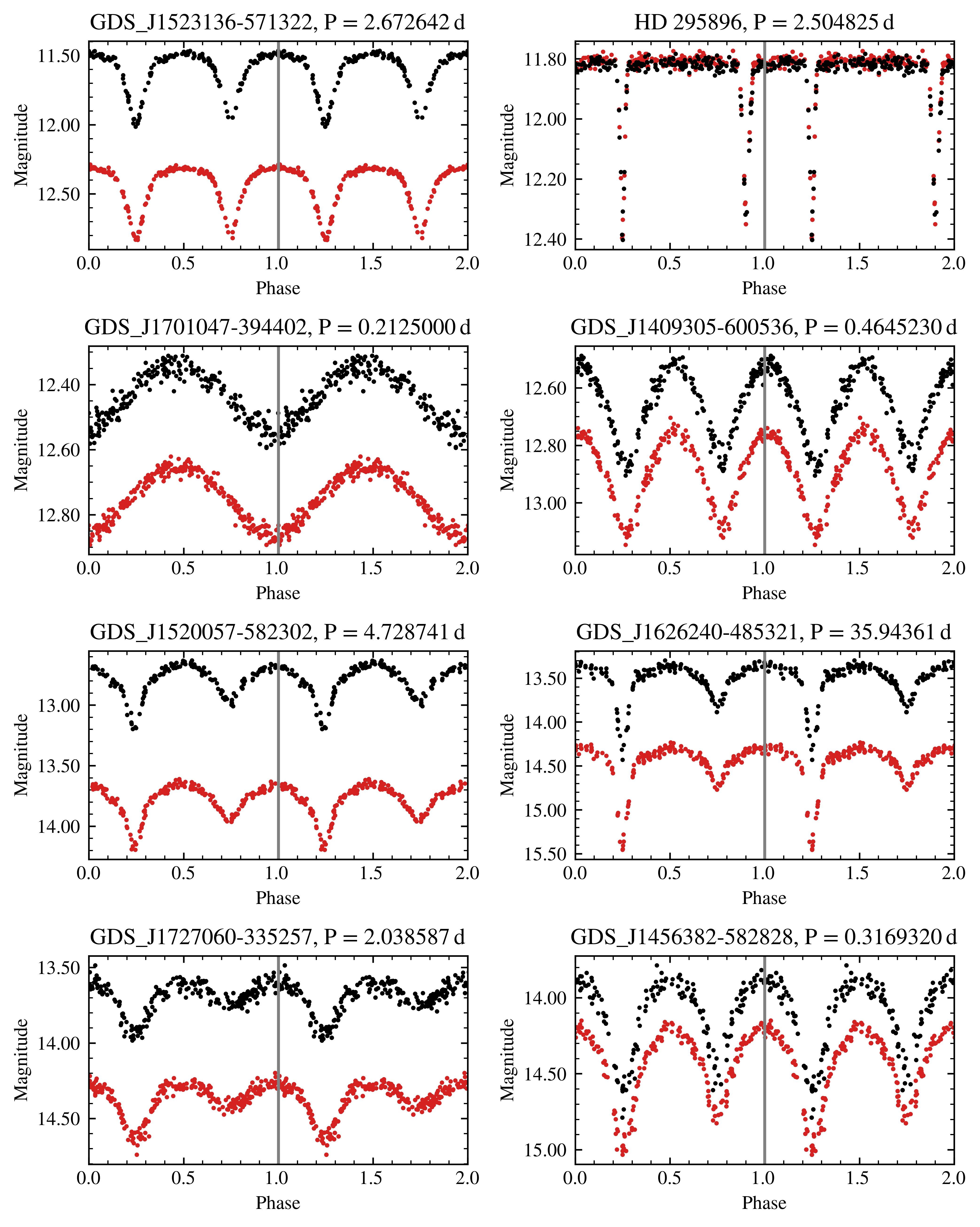}}
\addtocounter{figure}{-1}
\caption{\enspace Continued.\label{fig:LcsKnownVar2}}
\end{figure*}
\begin{figure*}[p]
\centerline{\includegraphics[width=1.0\textwidth]{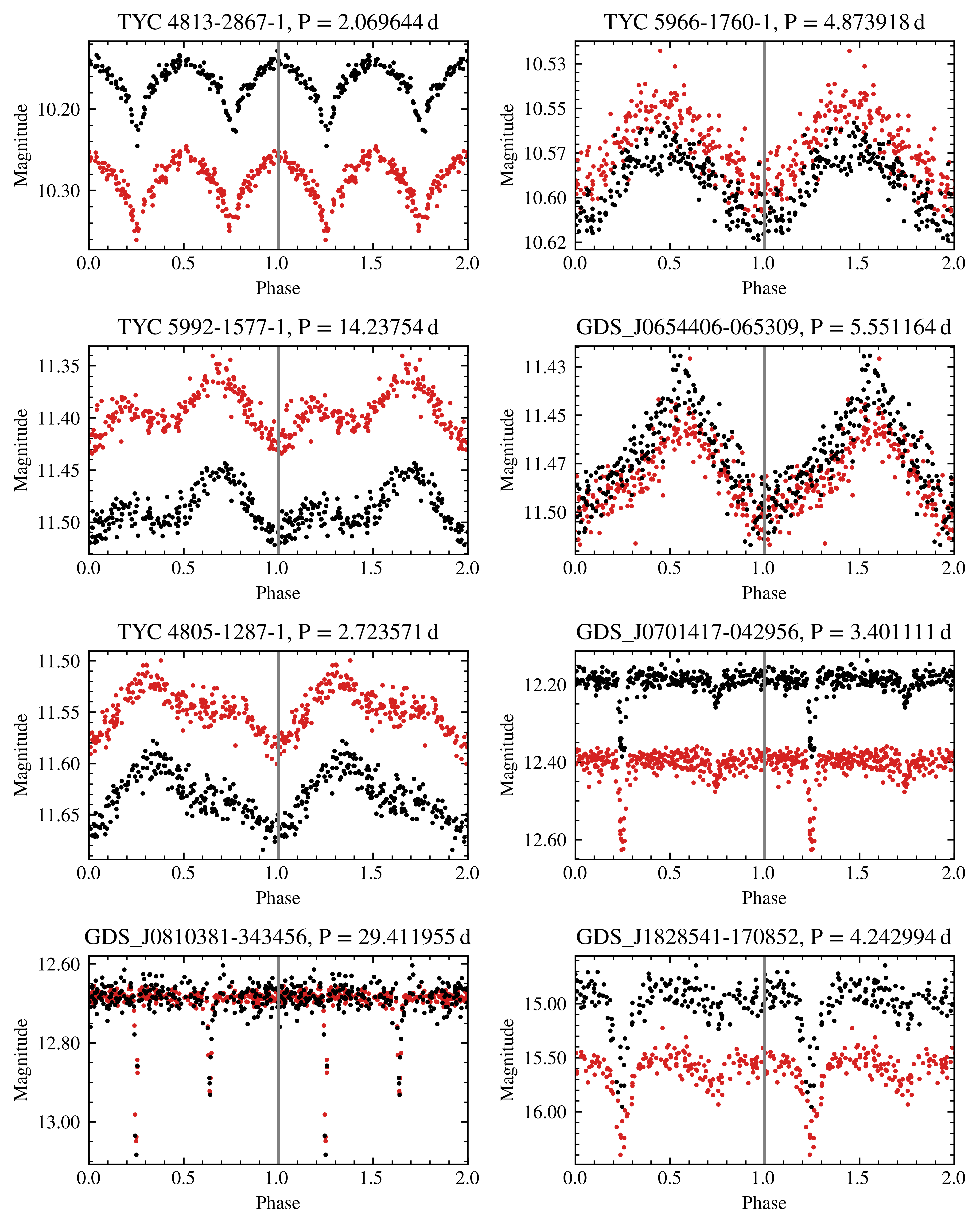}}
\caption{\enspace Folded $r'$ (red) and $i'$ (black) light curves for a selection of stars whose variability was previously unknown.\label{fig:LcsUnknownVar}}
\end{figure*}
\end{document}